\documentclass[aps,pra,twocolumn,superscriptaddress,showpacs,amsmath,amssymb]{revtex4}

\usepackage{graphicx}
\usepackage{epsfig}
\usepackage{bm}
\usepackage{amssymb}
\usepackage{hyperref}

\begin{document}

\title{Spatio-temporal vortex beams and angular momentum}

\author{Konstantin Y. Bliokh}
\affiliation{Advanced Science Institute, RIKEN, Wako-shi, Saitama 351-0198, Japan}
\affiliation{A. Usikov Institute of Radiophysics and Electronics, NASU, Kharkov 61085, Ukraine}

\author{Franco Nori}
\affiliation{Advanced Science Institute, RIKEN, Wako-shi, Saitama 351-0198, Japan}
\affiliation{Physics Department, University of Michigan, Ann Arbor, Michigan 48109-1040, USA}

\begin{abstract}
We present a \textit{space-time} generalization of the known spatial (monochromatic) wave vortex beams carrying intrinsic orbital angular momentum (OAM) along the propagation direction. Generic spatio-temporal vortex beams are polychromatic and can carry intrinsic OAM at an \textit{arbitrary angle} to the mean momentum. Applying either (i) a transverse wave-vector shift or (ii) a Lorentz boost to a monochromatic Bessel beam, we construct a family of either (i) time-diffracting or (ii) non-diffracting spatio-temporal Bessel beams, which are exact solutions of the Klein-Gordon wave equations. The proposed spatio-temporal OAM states are able to describe either photon or electron vortex states (both relativistic and nonrelativistic), and can find applications in particle collisions, optics of moving media, quantum communications, and astrophysics.
\end{abstract}


\maketitle

\section{Introduction}

Wavefront dislocations (phase singularities or optical vortices) were introduced in 1974 in a seminal paper by Nye and Berry [1] which gave rise to the field of singular optics [2]. A non-singular wave locally looks like a segment of a plane wave with parallel wave fronts, well-defined phase, and current flowing in the direction of wave propagation. Nye and Berry demonstrated that, in general, the wavefronts in 3D space contain topological dislocation lines, akin to those in crystal lattices, where the phase becomes singular and currents coil around the lines forming \textit{optical vortices}. The dislocations are characterized by a quantized vortex strength, also called topological charge. One can distinguish \textit{screw} and \textit{edge} wavefront dislocations where the vortex lines are, respectively, \textit{parallel} and \textit{orthogonal} to the wave propagation direction. In general, mixed \textit{edge-screw} dislocations, with the vortex core \textit{tilted} with respect to the phase fronts, are possible. Importantly, only screw dislocations are generic in \textit{monochromatic} wave fields, while generic edge-screw dislocations require additional \textit{temporal} variations [1,2].

In 1992 Allen \textit{et al.} revealed a close relation between optical vortices and wave \textit{orbital angular momentum} (OAM) [3]. They showed that axially-symmetric higher-order optical beams in free space bear screw dislocations along their axes and possess intrinsic OAM directed along the beam axis. This OAM is quantized with its magnitude (in units of $\hbar$ per one photon) being equal to the vortex charge. Thus, monochromatic \textit{vortex beams} represent the OAM states of light, and nowadays they play an important role in classical and quantum optics [4]. Remarkably, the above phase-dislocation-, vortex-, and OAM- properties are generic for all types of linear waves, both classical and quantum, independently of their nature. In particular, vortex beams with OAM were recently described and generated in electron microscopes [5,6], as well as employed in acoustics [7].

For monochromatic vortex wave beams, the intrinsic OAM is \textit{collinear} to the momentum, with its projection on the beam axis (helicity) being quantized. In other words, such OAM behaves very similar to the \textit{spin} of a massless particle (even for nonrelativistic electrons [5]). Then, a natural question arises: ``\textit{Can a generic wave packet or beam carry a well-defined intrinsic OAM in an arbitrary direction, i.e., tilted with respect to the propagation direction?}''. It would seem that the direction of the intrinsic OAM is associated with the direction of the vortex line, and the tilted OAM should appear in states with tilted vortices, i.e., mixed \textit{edge-screw} wavefront dislocations. As we pointed out, such dislocations are generic for \textit{polychromatic} fields and our question essentially requires studying vortex beams and OAM in \textit{space-time} [8,9].

In this paper, we address the above question and extend the concepts of intrinsic OAM and vortex beams to polychromatic states in space-time. We analyze the scalar Klein-Gordon wave equation, so that its relativistic and nonrelativistic limits describe both massless optical fields and massive Schr\"{o}dinger particles, assuming that the polarization effects can be neglected. We show that the existence of spatio-temporal vortex beams with tilted OAM follows from the requirement of relativistic invariance and can be obtained via Lorentz transformations of the usual spatial vortex beams [9]. Hence, the spatio-temporal OAM wave states can naturally appear from moving sources emitting stationary (monochromatic) vortex states in their rest frames. As such, our results have implications in optics of moving media [10], quantum communications with satellites [11], collisions of high-energy particles with OAM [12], and astrophysical applications of OAM of light [13,14].

\section{Monochromatic Bessel beams}

We start with the Klein-Gordon wave equation in units $\hbar=c=1$:
\begin{equation}
\label{eqn:1}
\left({-\partial_t^2 + \bm{\nabla}^2 - m^2 }\right)\psi = 0~,
\end{equation}
where $\psi(t,\bf{r})$ is the scalar wave function. The plane-wave solutions of this equation are:
\begin{eqnarray}\label{eqn:2}
\psi  \propto \exp \left[ {i\left( { - \omega \,t + {\bf k} \cdot {\bf r}} \right)} \right] & \equiv & \exp \left( {ik^\mu  r_\mu  } \right)~,\nonumber\\
\omega^2 - k^2 \equiv k^\mu  k_\mu & = & m^2~,
\end{eqnarray}
with $k^\mu = \left({\omega ,{\bf k}} \right)$ and $r^\mu = \left({t ,{\bf r}} \right)$ being the standard four-vectors in the Minkowski space-time with signature $(-,+,+,+)$. The second equality in Eq. (2) is the dispersion relation which determines the \textit{mass hyperboloid} (or the light cone at $m=0$) -- a hypersurface in momentum $k^\mu$-space, where we only consider the positive-energy domain $\omega >0$. We deal with the Klein-Gordon equation because it possesses relativistic space-time symmetry and is able to describe both quantum massive particles of different energies and classical waves. (Indeed, for $k\ll m$ it can be reduced to the Schr\"{o}dinger equation, whereas in the relativistic limit $k\gg m$ it becomes the usual wave equation). Alongside the wave function $\psi(t,\bf{r})$, we will use its plane-wave (Fourier) spectrum $\tilde{\psi}(\omega,\bf{k})$, defined with the explicit delta-function of the dispersion (2):
\begin{equation}
\label{eqn:3}
\psi \left( {r^\mu} \right) \propto \int {\tilde \psi \left( {k^\mu} \right)} \,\delta\!\left( {k^\mu k_\mu - m^2} \right)e^{ik^\mu r_\mu} d^4 k^\mu~.
\end{equation}

The simplest monochromatic solutions of wave equations, which carry OAM, are \textit{Bessel beams} [15]. These represent eigenmodes of the $z$-components of the momentum, ${\bf \hat p} = - i\bm{\nabla}$, and OAM, ${\bf \hat L} = {\bf \hat r} \times {\bf \hat p}$, and can be constructed as a superposition of multiple plane waves (2) with fixed $\omega=\omega_0$ and $k_z=k_{z0}$. The fixed frequency implies that the wave vectors form an \textit{isofrequency sphere} $k = \sqrt {\omega_{0}^2 - m^2 } \equiv k_0$ in $\bf{k}$-space, whereas a fixed $k_z$ cuts a \textit{circle} $k_\bot = \sqrt{\omega_{0}^2 - m^2 - k_{z0}^2} \equiv k_{\bot 0}$ on this sphere, see Fig.~1a. Hereafter $\left({k_\bot,\phi,k_z}\right)$ denote the cylindrical coordinates in $\bf{k}$-space. The relative phases of plane waves can grow around the circle and form a \textit{vortex} $\exp \left( {i\ell \phi } \right)$ with phase increment $2\pi\ell$ around the loop, whereas the vortex charge $\ell = 0,\pm 1,\pm 2,...$ determines the order of the Bessel beam and its OAM. Thus, the plane-wave spectrum of the $\ell$th-order beam can be written as
\begin{equation}
\label{eqn:4}
\tilde\psi_\ell \left({\omega,{\bf k}}\right) \propto \delta\!\left({k - k_{0}}\right)\delta\!\left({k_\bot - k_{\bot 0}}\right) \exp\!\left({i\ell \phi}\right)~.
\end{equation}
The corresponding real-space field (3) yields
\begin{equation}
\label{eqn:5}
\psi_\ell \left({t,{\bf r}}\right) \propto J_{\left| \ell \right|}\!\left( {k_{\bot 0} r_\bot}\right) \exp \left[{i\left({\ell \varphi + k_{z0}z - \omega_{0}t}\right)} \right]~,
\end{equation}
where $J_n\!\left(\xi \right)$ is the Bessel function of the first kind, and $\left({r_\bot,\varphi,z}\right)$ are cylindrical coordinates in real space. In the vicinity of the beam axis, $k_{\bot 0} r_\bot \ll 1$, 
\[
\psi_\ell \propto \left[{x + i\,{\rm sgn}\!\left(\ell\right)y} \right]^{\left|\ell\right|} \exp\left[{i\left({k_{z0}z - \omega_{0}t} \right)} \right]~,
\]
which demonstrates a screw wavefront dislocation of strength $\ell$ on the axis [1,2].

\begin{figure}[t]
\includegraphics[width=\columnwidth, keepaspectratio]{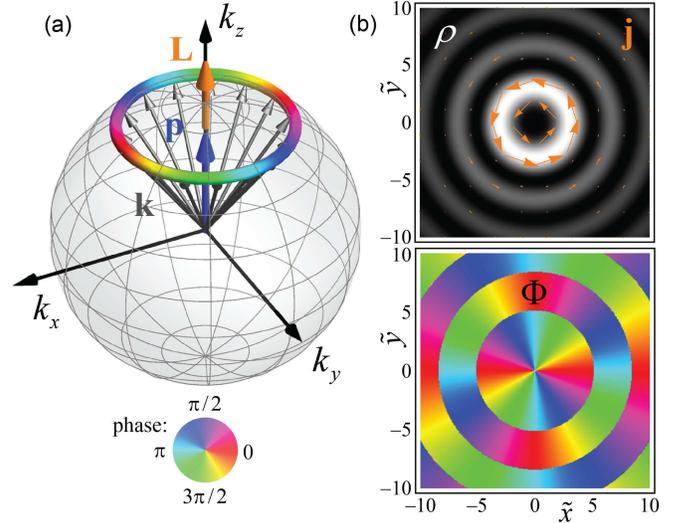}
\caption{(color online). \textbf{(a)} Plane-wave spectrum (4) of a monochromatic Bessel beam with $\ell=2$: a circle on the isofrequency sphere $k = \sqrt{\omega_0^2 - m^2}$ in $\bf{k}$-space. Relative phases of the constituent waves form a charge-2 optical vortex encoded by colors. \textbf{(b)} Transverse real-space distributions of the intensity $\rho = \left|\psi\right|^2$, current ${\bf j} = {\rm Im}\left({\psi^* \bm{\nabla} \psi}\right) = \rho \bm{\nabla} \Phi$, and phase $\Phi = \arg \psi$ for the Bessel beam (5) corresponding to Fig.~1a. The phase singularity on the beam axis represents a screw wavefront dislocation accompanied by a vortex of the current. The parameters used here are $\omega_0/m = 2$, $k_{\bot0}/m = 1$, $k_{z0}/m = \sqrt{2}$, and the dimensionless coordinates are $\tilde x = k_{\bot0} x$ and $\tilde y = k_{\bot0} y$. The normalized momentum and OAM, Eqs. (6) and (7), integrated over the visible area, yield ${\bf p} = k_{z0} {\bf e}_z$ and ${\bf L} = \ell {\bf e}_z$.} 
\label{fig1}
\end{figure}

Bessel beams (5) are non-diffracting, i.e., $\left| {\psi_\ell} \right|^2$ does not vary with $z$ and $t$ [15]. Evidently, these are eigenmodes of the operators $\hat p_z = -i\partial_z$ and $\hat L_z = -i\partial_\varphi$, with corresponding eigenvalues $k_z$ and $\ell$. Figure 1b shows the transverse spatial distributions of the intensity (probability density) $\rho = \left|\psi\right|^2$, current (momentum density) ${\bf j} = {\rm Im}\left({\psi^* \bm{\nabla} \psi}\right) = \rho \bm{\nabla} \Phi$, and phase $\Phi = \arg \psi$ for the beam (5). It is the screw phase dislocation at $r_\bot =0$ and the accompanying vortex current that produce a non-zero OAM of the field.

\begin{figure*}[tbh]
\centering\scalebox{0.5}{\includegraphics{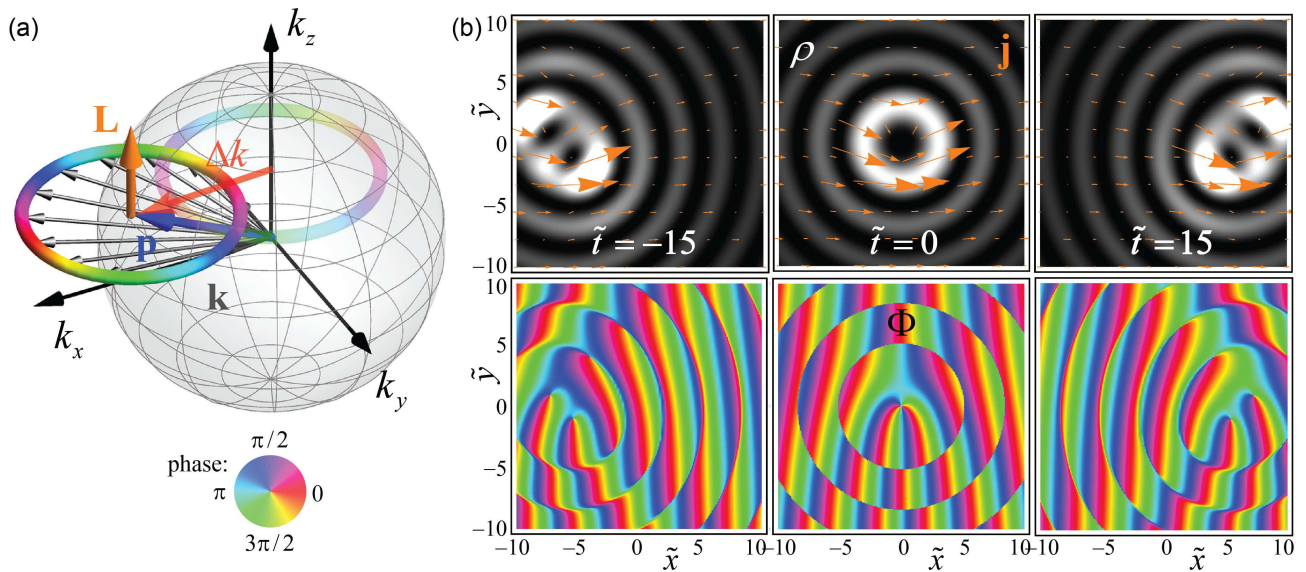}}
\caption{(color online). \textbf{(a)} Plane-wave spectrum of a polychromatic Bessel beam with $\ell=2$ and spectrum (9) shifted by $\Delta k \, {\bf e}_x$ in $\bf{k}$-space. Here the semitransparent circle represents the initial monochromatic beam, and the isofrequency sphere has radius $\sqrt{k_{z0}^2 + \Delta k^2}$. \textbf{(b)} Transverse real-space distributions of the intensity $\rho$, current $\bf{j}$, and phase $\Phi$ for the beam (a) at different times $t$. The beam moves in the positive $x$-direction and experiences time-diffraction deformations. The fork-like phase patterns in the beam center represent moving edge-screw wavefront dislocations. The parameters are $k_{z0}/m = 1$, $k_{\bot0}/m = 1.4$, $\Delta k/m = 2$, $\omega_c/m = 2.45$, and we use the dimensionless variables $\tilde x = k_{\bot0} x$, $\tilde y = k_{\bot0} y$, and $\tilde t = \omega_c t$. The normalized expectation values, Eqs.~(6) and (7), numerically integrated over the same area around the moving beam center yield Eqs.~(11) with a good accuracy: ${\bf R}_\bot \simeq \left({\Delta k/\omega_c}\right) t\, {\bf e}_x$, ${\bf p} \simeq k_{z0} {\bf e}_z + \Delta k \, {\bf e}_x$, 
${\bf L} \simeq \ell\, {\bf e}_z \nparallel {\bf p}$.} 
\end{figure*}

To characterize the angular momentum, it is important to also describe the expectation values of energy-momentum and coordinates of the beam [9,16]. The normalized (per one particle) expectation values of the energy, transverse coordinates, momentum, and OAM for the Bessel beam are:
\begin{eqnarray}
\label{eqn:6}
&& E = \frac{{\left\langle \psi  \right|i\partial _t \left| \psi  \right\rangle }}{{\left\langle \psi  \right|\left. \psi  \right\rangle }} = \omega_0,~
{\bf R}_ \bot   = \frac{{\left\langle \psi  \right|{\bf r}_ \bot  \left| \psi  \right\rangle }}{{\left\langle \psi  \right|\left. \psi  \right\rangle }} = {\bf 0},\nonumber\\
&& {\bf p} = \frac{{\left\langle \psi  \right| - i\bm{\nabla} \left| \psi  \right\rangle }}{{\left\langle \psi  \right|\left. \psi  \right\rangle }} = k_{z0} {\bf e}_z,~
{\bf L} = \frac{{\left\langle \psi  \right|{\bf \hat L}\left| \psi  \right\rangle }}{{\left\langle \psi  \right|\left. \psi  \right\rangle }} = \ell {\bf e}_z.
\end{eqnarray}
Hereafter ${\bf e}_a$ denotes the unit basis vector of the corresponding $a$-axis, whereas the subscript ``$\bot$'' indicates the transverse $(x,y)$-components of a vector. As expected, the OAM is collinear with the momentum, ${\bf L}\parallel {\bf p}$. The inner product in Eq.~(6) implies integration over the proper area of space. Formally, the Bessel beams are not localized, i.e., cannot be normalized in the whole space. However, this can be overcome by substituting delta-functions in the spectrum (4) with arbitrarily narrow Gaussian exponents, and the normalized expectation values (6) converge to finite values. An alternative form of Eqs.~(6) using the probability density $\rho$ and momentum density $\bf{j}$ distributions can be written as [17]
\begin{eqnarray}
\label{eqn:7}
&& {\bf R}_\bot = \frac{{\int{{\bf r}_\bot \rho} \,d^2{\bf r}_\bot}}{{\int{\rho } \,d^2{\bf r}_\bot}} = {\bf 0},~~
{\bf p} = \frac{{\int{\bf j}\,d^2{\bf r}_\bot}}{{\int{\rho}\,d^2{\bf r}_\bot}} = k_{z0} {\bf e}_z~,\nonumber\\
&& {\bf L} = \frac{{\int{\left( {{\bf r} \times {\bf j}} \right)}\,d^2{\bf r}_\bot}}{{\int{\rho}\,d^2{\bf r}_\bot}} = \ell\,{\bf e}_z.
\end{eqnarray}
This makes it clear that it is the circulation of the current $\bf{j}$ (shown in Fig.~1b) that produces the intrinsic OAM of the beam. Note that the proper probability density and momentum density for the relativistic Klein-Gordon equation differ from the ``naive'' $\rho$ and $\bf{j}$ used here [9,18]. However, this does not affect our consideration and will be discussed below in Section V.

In practice, the Bessel-beam spectrum (4) can be simplified and well approximated by a \textit{finite number} $N\gg 1$ of plane waves distributed over the spectral circle in $\bf{k}$-space (see Fig.~1a) [14]. In this manner,
\begin{eqnarray}
\label{eqn:8}
&& \psi_\ell \left({t,{\bf r}}\right) \propto \frac{1}{{\sqrt N }}\sum\limits_{n = 1}^N {\exp \left[ {i\left( { - \omega_n t + {\bf k}_n \cdot {\bf r}} + \ell \, \phi_n\right)} \right]},\nonumber\\
&& \omega_n = \omega_0,~~{\bf k}_n  = \left({k_{\bot 0}\cos\phi_n,k_{\bot0} \sin\phi_n,k_{z0}}\right),\nonumber\\
&& \phi_n = \frac{{2\pi}}{N}n.
\end{eqnarray}
The discretization (8) is important for interferometric applications [14] and approximates well the beams (5) in the restricted area around the axis: e.g., $k_{\bot 0} r_\bot \le 10$ for $N=20$. In Figure 1 and throughout this paper we use the superposition (8) of $N=30$ plane waves for numerical simulations of Bessel beams, because in some of the cases described below the beams cannot be characterized analytically. We verified numerically that the normalized energy, momentum, and OAM of such beam are in perfect agreement with Eqs. (6) and (7) when the integration is performed over the area $S = \left\{ {\left| {k_{\bot 0}x} \right| < 10,\left|{k_{\bot 0}y}\right| < 10} \right\}$.

\begin{figure*}[tbh]
\centering\scalebox{0.49}{\includegraphics{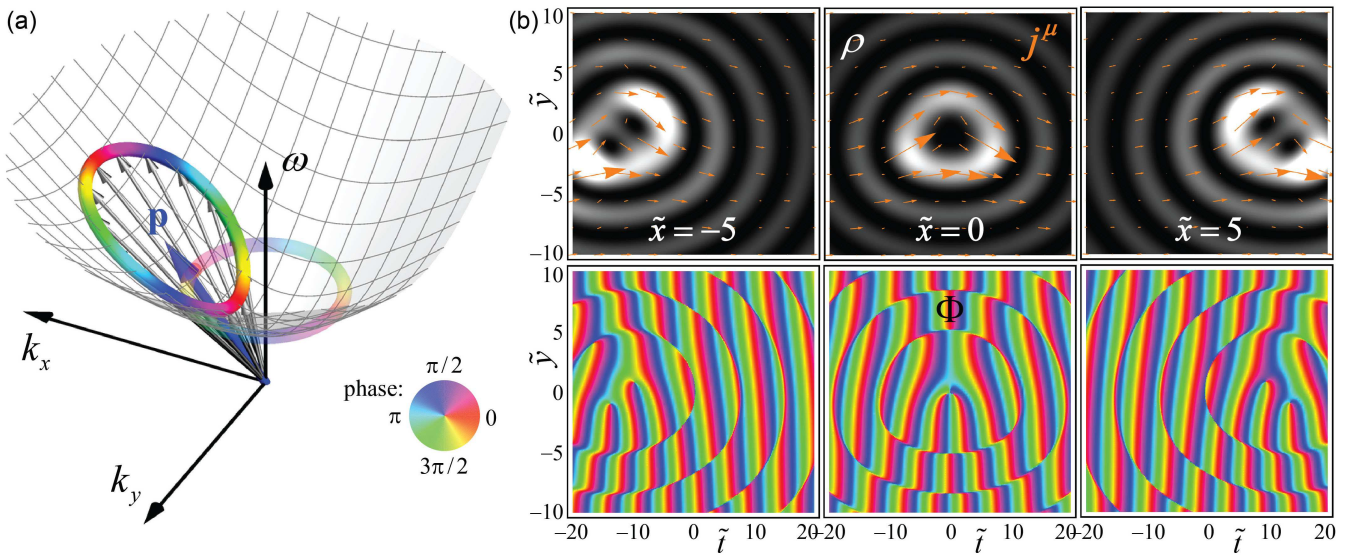}}
\caption{(color online). Spatio-temporal representation of the transversely-moving Bessel beam from Figure~2. \textbf{(a)} Here its spectrum is plotted on the mass hyperboloid in the $\left({\omega,k_x,k_y}\right)$-space where it is seen that it also forms a vortex loop if projected onto the $\left({\omega,k_y}\right)$ plane. \textbf{(b)} Real-space evolution shown here in the form of the distributions in the $(t,y)$ plane, for different values of $x$, cf. Fig.~2b. The corresponding components of the four-current $j^\mu$ are $(j^0,j_y)$, where $j^0 = - {{\rm Im}}\left({\psi^* \partial_t \psi}\right)$.} 
\end{figure*}

\section{Spatio-temporal Bessel beams with shifted spectrum}

From the above picture of the monochromatic Bessel beams, one can see that the expectation values of the OAM and momentum are collinear not by chance. Indeed, assuming a circular plane-wave distribution in momentum space (Fig.~1a), geometrically, $\bf{p}$ represents the radius-vector of the \textit{center of the circle}, whereas $\bf{L}$ points in the direction \textit{normal to the circle} and has a magnitude equal to the vortex charge $\ell$. Obviously, the collinearity ${\bf L}\parallel {\bf p}$ holds true for any circle on the isofrequency sphere, i.e., for any axially-symmetric monochromatic beam. However, as soon as we abandon the monochromaticity constraint, it is possible to construct a Bessel-beam-type solution with ${\bf L}\nparallel{\bf p}$. Indeed, \textit{any} circle in $\bf{k}$-space can serve as a plane-wave spectrum for the beam. The only fundamental constraint of the mass hyperboloid (2) can always be satisfied by choosing the corresponding frequencies $\omega\left({\bf k}\right) = \sqrt{m^2 + k^2}$. Thus, considering different spectral circles with vortex phases (see an example in Fig.~2a), we obtain different vortex solutions of the wave equation (1). It is easy to show that the center of the circle and the normal to the circle still represent the mean momentum and intrinsic OAM of the beam, respectively.

The simplest transformation shifting the spectral circle away from the isofrequency sphere is a uniform transverse shift (e.g., along the $x$-axis) of all the wave vectors in the Bessel-beam spectrum (Fig.~2a):
\begin{equation}
\label{eqn:9}
{\bf k} \to {\bf k} + \Delta k \, {\bf e}_x,~~\omega_0 \to \omega\!\left({\bf k}\right) = \sqrt{\omega_0^2 + \Delta k^2 + 2k_x \Delta k}.
\end{equation}
Solutions with spectrum (9) are polychromatic because of the $\omega(k_x)$ dependence (here $k_x$ is the wave vector component of the original monochromatic beam). Remarkably, the beam spectrum (9) represents a circle with an azimuthal phase gradient not only when projected onto the $(k_x,k_y)$ plane, but also a similar vortex loop around $\omega = \omega_c \equiv \sqrt{\omega_0^2 + \Delta k^2}$ in the $(\omega,k_y)$ plane. Thus, this is a vortex loop on the mass hyperboloid in $(\omega,\bf{k})$ space, which demonstrates its \textit{spatio-temporal} nature, see Fig.~3a.

\begin{figure*}[tbh]
\centering\scalebox{0.50}{\includegraphics{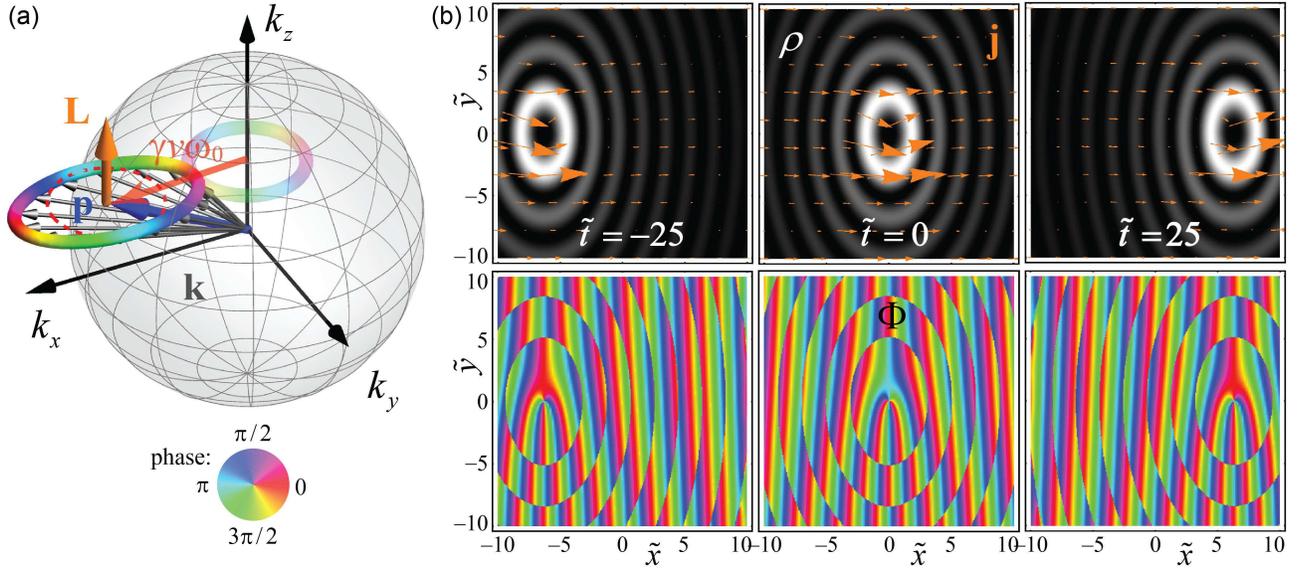}}
\caption{(color online). \textbf{(a)} Plane-wave spectrum (15) of a Lorentz-boosted Bessel beam with $\ell = 2$ in  $\bf{k}$-space. Here the semitransparent circle represents the initial monochromatic beam, whereas the isofrequency sphere has radius $\sqrt{k_{z0}^2 + \left({\gamma v \omega_0}\right)^2}$. \textbf{(b)} Transverse real-space distributions of the probability density $\rho$, current $\bf{j}$, and phase $\Phi$ for the beam (14) corresponding to \textbf{(a)}, at different times $t$. The beam moves uniformly in the positive $x$-direction without diffraction. The fork-like phase pattern represents a moving edge-screw wavefront dislocation. The parameters are $\omega_0/m=2$, $\kappa_0/m=1$, $\Delta k/m=2$, $\nu=0.8$ ($\gamma=5/3$), and we use the dimensionless variables $\tilde x =k_{\bot0}x$, $\tilde y =k_{\bot0}y$, and $\tilde t =\omega_c t$. The numerically-calculated normalized expectation values, Eqs.~(6) and (7), correspond to Eq.~(18) with good accuracy.} 
\label{fig4}
\end{figure*}

Formally, the wave function of the beam can be written as a Fourier integral (3) with the delta-function spectrum (4) shifted by the transformation (9). However, such integral cannot be evaluated analytically in the general case. To understand properties of the beams with a shifted spectrum, we performed numerical simulations using the discretization (8) with parameters corresponding to Eq.~(9):
\begin{eqnarray}
\label{eqn:10}
&& \omega_n  = \sqrt{\omega_0^2 + \Delta k^2 + 2k_{\bot 0} \Delta k\cos\phi_n}~,\nonumber\\
&& {\bf k}_n = \left({\Delta k + k_{\bot 0} \cos\phi_n, k_{\bot0} \sin\phi_n, k_{z0}}\right)~.
\end{eqnarray}
The results are presented in Figure~2b. We see a Bessel-beam solution which is still homogeneous in the $z$-direction ($k_z=k_{z0}$) but evolving in time. At $t=0$, the intensity distribution represents the Bessel function: $\rho\left({{\bf r},t = 0}\right) \propto \left[{J_{\left|\ell\right|}\!\left({k_{\bot 0}r_\bot}\right)}\right]^2$, but the current and phase distributions differ significantly as compared with the monochromatic beam (5) (Fig.~1b). Indeed, there is a net transverse current flowing in the $x$-direction (because of the mean $\left\langle {k_x } \right\rangle = \Delta k$), whereas the phase dislocation in the beam centre becomes of mixed \textit{edge-screw} type. This is seen from the fork-like pattern of the phase fronts instead of the radial pattern characteristic of screw dislocations. Calculating the wave-field distributions at $t\neq 0$, we see that the beam \textit{moves in the transverse $x$-direction} and experiences shape distortions. The deformations represent \textit{diffraction in time} caused by the accumulated phase difference between waves with different frequencies (10). In particular, Fig.~2b shows that this diffraction breaks the $\ell$th-order phase dislocation at $t=0$ into $|\ell|$ basic dislocations of   strengths at $t\neq 0$.
This can be regarded as a temporal manifestation of the general instability of higher-order vortices with respect to perturbations [1,2,19].

The moving edge-screw dislocation reveals a spatio-temporal character of the shifted-spectrum vortex beam. To illustrate this, we plot the beam density, current, and phase distributions not only in the $(x,y)$ plane for different values of $t$ (Fig.~2b), but also in the $(t,y)$ plane for different values of $x$, in Fig.~3b. One can see quite similar ``moving'' beams with edge-screw phase dislocation and circulating probability current in both the $(x,y)$ and $(t,y)$ planes. Thus, in general, this is a vortex ``hyper-beam'' in Minkowski $(t,\bf{r})$ space-time.

The transverse motion of the beam implies a non-zero transverse momentum, $p_x\neq 0$, whereas the OAM is still determined by the $z$-direction of the vortex axis (which also moves along $x$ with time). Calculating the expectation values (6) in the momentum (Fourier) representation and assuming paraxiality of the beam, $k_{\bot 0} \ll \sqrt{k_{z0}^2 + \Delta k^2}$, we obtain
\begin{eqnarray}
\label{eqn:11}
E & \simeq & \omega _c = \sqrt {\omega _0^2  + \Delta k^2 }~,~~{\bf R}_\bot \simeq \frac{{\Delta k}}{{\omega_c}}\, t \, {\bf e}_x~,\nonumber\\
{\bf p} & \simeq & k_{z0} {\bf e}_z  + \Delta k\,{\bf e}_x~,~~{\bf L} \simeq \ell \, {\bf e}_z~.
\end{eqnarray}
Here the beam centroid moves in the transverse direction according to the free-space equation of motion: $d{\bf R}_\bot /dt = {\bf p}_\bot /E$. Numerical calculations of the expectation values in the representation (7), performed using the same area around the instantaneous transverse beam centroid ${\bf R}_\bot$, showed good agreement with Eq.~(11) even for the non-paraxial parameters used in Fig.~2. Thus, Figures~2 and 3, together with Eqs.~(11), demonstrate the existence of spatio-temporal vortex beams with non-collinear momentum and intrinsic OAM: ${\bf L}\nparallel{\bf p}$. 

\section{Lorentz-boosted spatio-temporal Bessel beams}

It turns out that it is possible to construct a family of spatio-temporal beams which are free of the temporal diffraction and allow a simple analytic description. Note that Lorentz transformation to a moving reference frame, $r^\mu \to \hat\Lambda^{\mu}_{~\,\nu}\!\left({\bf v}\right)r^\nu$ (${\bf v}$ is velocity), provides a boost of the four-momentum in the spectrum, $k^\mu \to \hat\Lambda^{\mu}_{~\,\nu}\!\left(-{\bf v}\right)k^\nu$, and can generate polychromatic moving solutions. Since the Klein-Gordon wave equation is Lorentz-invariant, the Lorentz-boosted Bessel beams (4) and (5) will also represent exact solutions of Eq.~(1):
\begin{equation}
\label{eqn:12}
\psi^{\prime}_{\ell} \left({r^\nu}\right) \equiv \psi_\ell\!\left[{\hat\Lambda^{\mu}_{~\,\nu}\!\left({\bf v}\right)r^\nu}\right],~
\tilde \psi^\prime_\ell \left({k^\nu}\right) \equiv \tilde\psi_\ell\!\left[{\hat\Lambda^{\mu}_{~\,\nu}\!\left({-{\bf v}}\right)k^\nu} \right].
\end{equation}
Equations (12), with Eqs.~(4) and (5), offer a new family of \textit{moving} vortex beams parameterized by the velocity $\bf{v}$.

\begin{figure*}[tbh]
\centering\scalebox{0.49}{\includegraphics{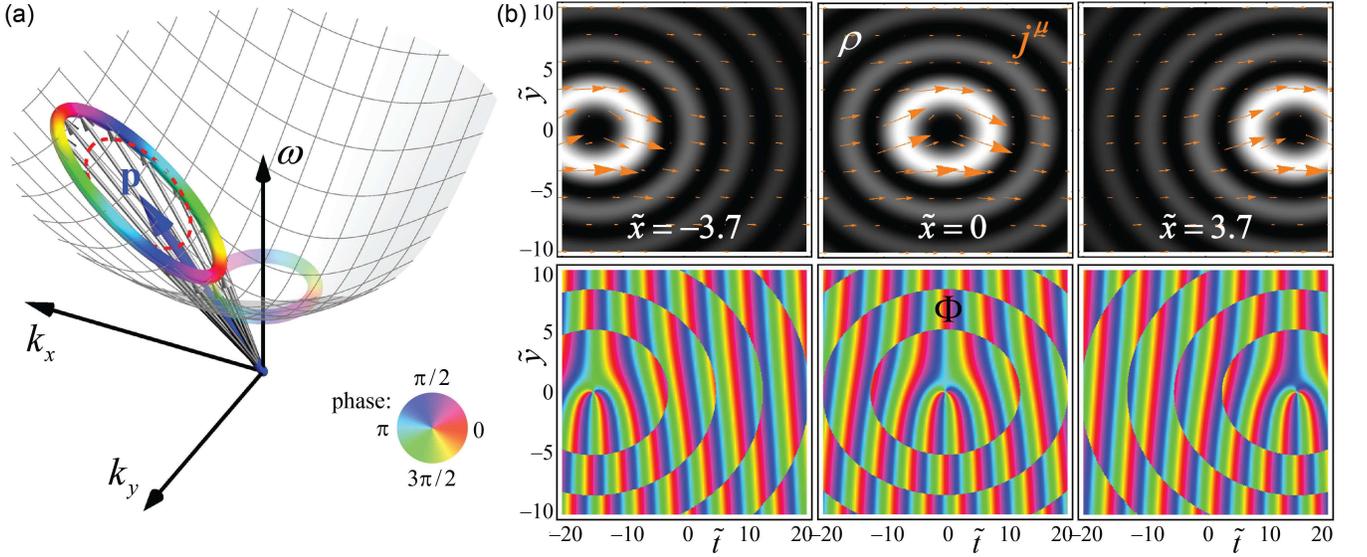}}
\caption{(color online). Spatio-temporal representation of the transversely-moving Bessel beam from Figure 4. \textbf{(a)} Here its spectrum is plotted on the mass hyperboloid in the $\left({\omega,k_x,k_y}\right)$-space where it is seen that it also forms an elliptical vortex loop if projected onto the $(\omega,k_y)$ plane. \textbf{(b)} The real-space evolution of the beam is shown here in the form of the density, current, and phase distributions in the $(t,y)$ plane for different values of $x$, cf. Fig.~4b. The corresponding components of the four-current $j^\mu$ are $(j^0,j_y)$, where $j^0 = - {{\rm Im}}\left({\psi^* \partial_t \psi}\right)$.} 
\label{fig5}
\end{figure*}

It is easy to see that the Lorentz boosts in the longitudinal $z$-direction keep the monochromatic Bessel-beam family (4) and (5) invariant and only transform their parameters $\omega_0$, $k_{z0}$, and $k_{\bot0}$. Therefore, we consider only the non-trivial case of the \textit{transverse} boost, say, in the $x$-direction. Assuming $\bf{v}=v\,\bf{e}_x$, we have
\begin{equation}
\label{eqn:13}
\hat \Lambda^{\mu}_{~\,\nu}\!\left( {\bf v} \right) = \left( {\begin{array}{*{20}c}
   \gamma  & { - v\gamma } & 0 & 0  \\
   { - v\gamma } & \gamma  & 0 & 0  \\
   0 & 0 & 1 & 0  \\
   0 & 0 & 0 & 1  \\
\end{array}} \right),
\end{equation}
with $\gamma = 1/\sqrt{1 - v^2}$ being the Lorentz factor. Explicitly, the moving Bessel beam (12) with Eq.~(13) is obtained via the substitution $t \to \gamma\left({t - vx}\right)$, $x \to \gamma \left( {x - vt} \right)$ in Eq.~(5), and it takes the form
\begin{eqnarray}
\label{eqn:14}
&& \psi^\prime_\ell\!\left({t,{\bf r}}\right) \propto J_{\left|\ell\right|}\!\left({k_{ \bot 0} r^\prime_\bot} \right)\exp\!\left[\,{i\left( {\ell \varphi^\prime + k_{z0}z - \omega_{0}t^\prime}\right)}\right],\nonumber\\
&& r^\prime_\bot = \sqrt{\gamma^2 \left({x - vt}\right)^2 + y^2},~~\varphi^\prime = \tan^{-1}\!\left({\gamma \frac{{x - vt}}{y}} \right),\nonumber\\
&& t^\prime = \gamma\left({t - vx}\right).
\end{eqnarray}
At the same time, the Lorentz transformation (12) and (13) of the spectral characteristics yields [cf. Eq.~(9)]:
\begin{equation}
\label{eqn:15}
k_x \to \gamma\left({k_x + v \omega_0}\right),~~
\omega_0 \to \omega\left({\bf k}\right) = \gamma\left({\omega_0 + v k_x}\right).
\end{equation}
In $\bf{k}$-space, this represents a shift of the spectral circle by the distance $\Delta k_x = \gamma v\omega_0$ and a stretch with factor $\gamma$ along the $k_x$-dimension, see Fig.~4a. This stretching of the spectrum ensures the linearity of the $\omega(k_x)$-dependence which causes \textit{no diffraction} in time. Plotting the Lorentz-boosted spectrum (15) in the $(\omega,\bf{k})$ space, one can see an elliptic planar loop on the mass hyperboloid, Fig.~5a. One can consider this as a vortex loop both when projected onto the   plane and also on the $(k_x,k_y)$ plane, which demonstrates the spatio-temporal character of this vortex. The spectral discretization (8) for the Lorentz-boosted beam acquires the following form according to Eq.~(15):
\begin{eqnarray}
\label{eqn:16}
&& \omega_n = \gamma\!\left({\omega_0 + vk_{\bot 0} \cos\phi_n}\right),\nonumber\\
&& {\bf k}_n = \left({\gamma\!\left({k_{ \bot 0} \cos\phi_n + v\omega_0} \right),k_{\bot 0}\sin\phi_n,k_{z0}} \right).
\end{eqnarray}
%

\begin{figure*}[tbh]
\includegraphics[width=14cm, keepaspectratio]{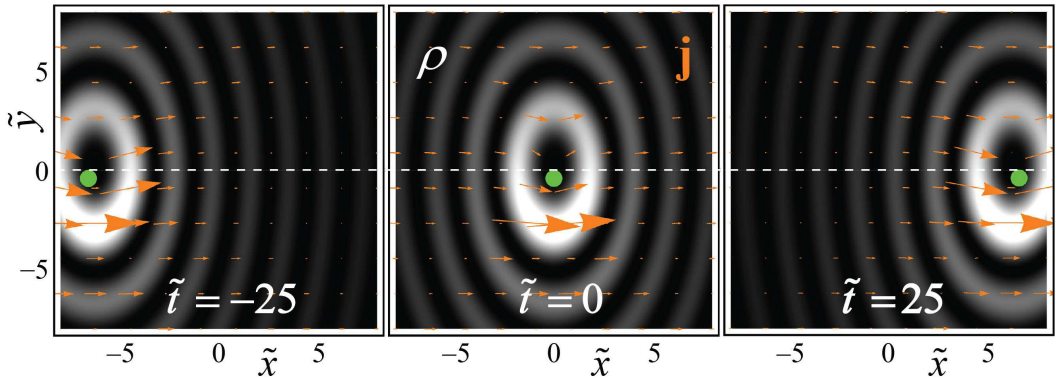}
\caption{(color online). Same evolution of the Lorentz-boosted Bessel beam (14) as in Fig.~4, but here with the scaling $\tilde\psi \to \sqrt\omega \tilde\psi$ of the spectral Fourier amplitudes. Thus, $\rho  = \left| \psi \right|^2$ becomes the true relativistic probability density, and ${\bf j} = {\rm Im}\left({\psi^*\bm{\nabla}\psi}\right)$ becomes the energy current (momentum density). This causes the relativistic Hall effect [9]: a tiny $y$-deformation and subwavelength shift (19) of the beam which is seen here owing to the nonparaxial character (small radius) of the beam. The dot indicates the centre of gravity of the beam according to Eqs. (18) and (19): ${\bf R}_\bot \simeq \left({vt,-{v\ell}/{2\omega_0}}\right)$, and the small $x$-deviations of the dot are due to the non-paraxial corrections.} 
\label{fig6}
\end{figure*}

Figure 4b shows the real-space distributions of the intensity, current, and phase in the Lorentz-boosted Bessel beam (14). This beam moves in the transverse $x$-direction with velocity $\bf{v}=v \bf{e}_x$ and has an elliptical profile (stretching of $k_x$ yields the Lorentz contraction of the $x$-dimension with factor $\gamma$). The wave function (14) near the beam axis, $k_{\bot 0} r^\prime_\bot \ll 1$, is approximated by
\begin{eqnarray}
\label{eqn:17}
\psi^\prime_\ell & \propto & \left[\,{\gamma\!\left({x - vt}\right) + i\,{\rm sgn}\!\left(\ell\right)y}\, \right]^{\left| \ell  \right|}\nonumber\\ 
& \times & \exp\!\left[\,{i\left({k_{z0}z + \omega_{0}\gamma\left({vx - t}\right)}\right)}\right]~.
\end{eqnarray}
It has the form of a mixed edge-screw dislocation [1,2], i.e., a spatio-temporal vortex in the $(t,x,y)$ space, with its singularity line $\left\{{y = 0,x = vt}\right\}$ being parallel to the $z$-axis and moving along the $x$-axis. Figure 5b illustrates the spatio-temporal character of the Lorentz-boosted beam (14) by showing its evolution in the $(t,y)$ plane for different values of $x$. One can see that the $x$-moving vortex appears in the $(t,y)$ plane entirely similar to the $t$-moving vortex in the $(x,y)$ plane.

Similar to the example discussed in the previous Section, the transverse current and motion of the Lorentz-boosted beam implies a tilted momentum of the field, $p_x\neq 0$, whereas the OAM is still determined by the $z$-direction of the vortex axis. Calculating the expectation values (6) and (7) and assuming paraxiality of the beam, $\gamma k_{\bot 0} \ll \sqrt{k_{z0}^2 + \left({\gamma v \omega_0}\right)^2}$, we obtain:
\begin{eqnarray}
\label{eqn:18}
&& E \simeq \omega_c = \sqrt {\omega_0^2 + \left({\gamma v \omega_0} \right)^2} = \gamma\,\omega_0,~~
{\bf R}_\bot \simeq vt\,{\bf e}_x,\nonumber\\ 
&& {\bf p} \simeq k_{z0} {\bf e}_z + \gamma v \omega_0 {\bf e}_x,~~
{\bf L} \simeq \frac{{\gamma + \gamma^{-1}}}{2}\ell\,{\bf e}_z.
\end{eqnarray}
Here the values of the transverse coordinates and momentum correspond to the linear motion of the beam, $d{\bf R}_\bot/dt = {\bf p}_\bot/E$, whereas the transformation of the OAM, which acquires the factor $\left( {\gamma + \gamma^{-1}}\right)/2$, owes its origin to the elliptical relativistic deformation of the beam [9,20]. Indeed, the OAM is given by the cross-product of the position and momentum, $\hat L_z  = \hat x \hat p_y - \hat y \hat p_x$, and both summands $\hat x \hat p_y$ and $- \hat y \hat p_x$ provide equal contributions of $L_z/2$ to the intrinsic OAM of the cylindrical beam (5). In the transversely-moving beam (14) and (15), the Lorentz transformation results in the contraction of the coordinate $x$ and expansion of the momentum component $k_x$ (Fig.~3), and the above two summands acquire factors $\gamma^{-1}$ and $\gamma$, respectively. Thus, Eqs.~(12)--(18) and Figures 4 and 5 demonstrate the existence of \textit{spatio-temporal non-diffracting Bessel beams} with non-collinear momentum and intrinsic OAM, ${\bf L}\nparallel{\bf p}$. 

\section{Relativistic Hall effect}

It should be noticed that the last equation (18) contradicts the Lorentz transformation of the angular momentum $\bf{L}$ of a point particle. Indeed, the Lorentz transformation of the original OAM ${\bf L} = \ell\, {\bf e}_z$ should yield ${\bf L}^\prime = \gamma \ell \, {\bf e}_z$ after the boost (13). This apparent paradox is considered in detail in [9]. It is resolved when one takes into account the ``relativistic Hall effect'', i.e., the transverse $y$-deformation and shift of the OAM-carrying object observed in an $x$-moving frame. It turns out that the geometric center of such object undergoes a transverse shift of its centroid:
\begin{equation}
\label{eqn:19}
Y \simeq  - \frac{{v}}{{2E_0 }}\,L =  - \frac{{v \ell }}{{2\omega_0}}.
\end{equation}
Together with the momentum component $p_x \simeq \gamma v \omega_0$, Eq.~(18), the shift (19) generates \textit{extrinsic OAM} $L^{\rm (ext)}  =  - Y p_x  = \gamma v^2 \ell /2$. Adding it to the intrinsic OAM (18), we obtain 
\begin{equation}
\label{eqn:20}
L^{({\rm int})} + L^{\rm (ext)} \simeq \frac{{\gamma + \gamma^{-1}}}{2}\ell + \frac{{\gamma v^2}}{2}\ell  = \gamma \ell~,
\end{equation}
in agreement with the Lorentz transformation.

We did not observe the shift (19) in Figures 4 and 5, because we used the ``naive'' density $\rho  = \left| \psi \right|^2$ and current ${\bf j} = {\rm Im}\left({\psi^*\bm{\nabla}\psi}\right)$. Indeed, $\rho$ does not describe the density of particles in the case of the relativistic Klein-Gordon equation (1) because the volume integral $\int{\rho} dV$ is not Lorentz-invariant. The proper particle density and current are given by [18]
\begin{equation}
\label{eqn:21}
\rho_P = -{\rm Im}\left({\psi^*\partial_t\psi}\right),~~
{\bf j}_P = {\rm Im}\left({\psi^*\bm{\nabla}\psi}\right),
\end{equation}
whereas the energy density and current (momentum density) are:
\begin{eqnarray}
\label{eqn:22}
&& \rho_E = \frac{1}{2}\left[{\left|{\partial_t \psi}\right|^2 + \left|{\bm{\nabla}\psi}\right|^2 + \mu^2\left| \psi\right|^2} \right]~,\nonumber\\ 
&& {\bf j}_E  =  - {\rm Re}\!\left[ {\left( {\partial _t \psi } \right)^*\!\left( {\bm{\nabla} \psi } \right)} \right]~.
\end{eqnarray}
For plane waves, this yields simple $\omega$-scalings: $\rho_E = \omega^2 \rho$, ${\bf j}_E = \omega\,{\bf j}$, and $\rho_P = \omega\,\rho$, which make no difference for monochromatic beams in the rest frame. However, a Lorentz transformation to the moving frame affects these distributions via local variations of the frequency. In other words, one can use the ``naive'' density $\rho$ and current $\bf{j}$, instead of $\rho_P$ and $\bf{j}_E$ which should be used in Eqs.~(6), but the amplitudes of the Fourier spectral components should be scaled as $\tilde\psi \to \sqrt\omega\, \tilde\psi$.

In Figure 6 we plot the spatio-temporal evolution of the Lorentz-boosted beam with the Fourier amplitude scaling $\tilde\psi \to \sqrt\omega\, \tilde\psi$ taking into account the relativistic Hall effect [9]. The $y$-deformation and shift (19) are clearly seen owing to the chosen nonparaxial relativistic parameters of the beam. However, for paraxial or nonrelativistic beams the tiny shift (19) (less than a fraction of the wavelength) is practically imperceptible. Moreover, we emphasize that both the spatio-temporal beams with scaling $\tilde\psi \to \sqrt\omega\, \tilde\psi$ and those without it are exact solutions of the Klein-Gordon wave equation and both can exist.

\section{Concluding remarks}

We have demonstrated the existence of spatio-temporal vortex wave beams in free space. Such beams move uniformly in the transverse direction and carry intrinsic orbital angular momentum non-collinear to their mean momentum. Furthermore, the spatio-temporal vortex around the beam axis appears as a moving edge-screw dislocation in both the spatial and space-time cross-sections of the beam. The spatio-temporal vortex beams are polychromatic and can naturally appear in problems with non-stationary (moving) sources or media. Localized wave-packets with spatio-temporal vortices and tilted intrinsic OAM can readily be constructed by considering slightly delocalized (e.g., Gaussian) Fourier spectra instead of the delta-functions considered here. It should be emphasized that our analysis is valid for both massless and massive fields, in both relativistic and non-relativistic cases.

Quite naturally, a complete relativistic family of spatio-temporal Bessel beams is constructed via the Lorentz transformations of the spatial (monochromatic) Bessel beams. This means that even the stationary OAM states of light or quantum particles will be seen as spatio-temporal states in the case of a transversely-moving observer or source. In this manner, the deformation of the phase pattern of a moving vortex [cf. Fig.~2 or 4 with Fig.~1] offers a sensitive interferometric tool detecting relativistic effects even at non-relativistic velocities. Indeed, purely relativistic deformations of the phase fronts (intimately related to the Lorentz transformation of time [9]) become significant for velocities $v$ comparable with $c^2 /\omega \,r\sim c^2 k_{\bot0}/\omega$, where $r\sim k_{\bot0}^{-1}$ is the beam radius. For paraxial beams with $\theta \equiv k_{ \bot 0}c/\omega \ll 1$, this allows the observation of strong deformations at speeds $v\sim\theta c \ll c$. This is explained by the fact that the transverse phase velocity of the motion of the phase fronts is much larger than the speed of light, $\omega /k_{\bot 0} = \theta^{-1}c \gg c$.

Spatio-temporal OAM states of light or particles can appear in a variety of systems involving moving frames and sources. In particular, they can be created upon scattering by moving objects, emitted by satellites doing quantum communications [11] or by natural astrophysical sources [13,14], and produced in collisions of high-energy particles [12]. Furthermore, transversely-moving vortex solutions can be important in two-dimensional wave fields, such as surface plasmon-polaritons [21]. While we have discussed the simplest \textit{vortex-beam} configurations, belonging to the Bessel-beam family, note that optical vortex \textit{lattices} can appear from the generic interference of only three monochromatic plane waves [22]. In a similar manner, an infinite lattice of spatio-temporal vortices can appear from the generic interference of three plane waves with different wave vectors and frequencies. This offers a simple way of generating spatio-temporal vortices in an optical laboratory.

In this work we only considered scalar waves and orbital angular momentum. For vector (e.g., electromagnetic) waves carrying \textit{spin} angular momentum, the relativistic transformation properties can be more complicated. Indeed, the spin angular momentum is produced by a circulating \textit{spin current}, which, however, does not transport energy [17,23]. This current is unrelated to phase gradients and have specific properties distinct from those of the regular orbital current. Although one could expect that the integral angular momentum of the beam is transformed in the same way independently of its spin or orbital nature, the internal deformations of the beam can be different. For instance, it is known that spin has a purely \textit{intrinsic} nature, and cannot have extrinsic contributions produced by the relativistic Hall effect, Eq.~(20). Relativistic transformations of vector beams carrying both spin and orbital angular momenta will be considered in detail elsewhere.

\section*{ACKNOWLEDGEMENTS}
We acknowledge fruitful discussions with I.P. Ivanov, correspondence with A.P. Sukhorukov, and support from the European Commission (Marie Curie Action), ARO, JSPS-RFBR contract No. 12-02-92100, Grant-in-Aid for Scientific Research (S), MEXT Kakenhi on Quantum Cybernetics, and the JSPS via its FIRST program.


\end{document}